\documentclass[
reprint,  floatfix,
amsmath,amssymb,aps,showpacs,superscriptaddress,dvipdfm   %
nofootinbib, showkeys
]{revtex4-1}
\usepackage{caption}
\usepackage{amssymb}
\usepackage{graphicx}
\usepackage{booktabs}
\usepackage{dcolumn}
\usepackage{bm}
\usepackage{mathrsfs}
\usepackage{mathtools}
\usepackage{amsmath}
\usepackage{epstopdf}
\usepackage{units}
\usepackage{mathtools}
\usepackage{ragged2e}
\usepackage[mathlines]{lineno}
\usepackage{dsfont}
\usepackage{upgreek}
\usepackage{subfig}
\usepackage{tikz}
\begin{document}
\preprint{APS/123-QED}

\title{Scaling properties of noise-induced switching in a bistable tunnel diode circuit}

\author{Stephen W. Teitsworth}
\affiliation{Duke University, Department of Physics, Box 90305
Durham, NC 27708-0305}

\author{Matthew E. Olson}
\affiliation{Duke University, Department of Physics, Box 90305
Durham, NC 27708-0305}

\author{Yuriy Bomze}
\affiliation{Duke University, Department of Physics, Box 90305
Durham, NC 27708-0305}

\date{\today}

\begin{abstract}
Noise-induced switching between coexisting metastable states occurs in a wide range of far-from-equilibrium systems including micro-mechanical oscillators, epidemiological and climate change models, and nonlinear electronic transport in tunneling structures such as semiconductor superlattices and tunnel diodes.  In the case of tunnel diode circuits, noise-induced switching behavior is associated with negative differential resistance in the static current-voltage characteristics and bistability, i.e., the existence of two macroscopic current states for a given applied voltage.  Noise effects are particularly strong near the onset and offset of bistable current behavior, corresponding to bifurcation points in the associated dynamical system.  In this paper, we show that the tunnel diode system provides an excellent experimental platform for the precision measurement of scaling properties of mean switching times versus applied voltage near bifurcation points.  More specifically, experimental data confirm that the mean switching time scales logarithmically as the 3/2 power of voltage difference over an exceptionally wide range of time scales and noise intensities.

\end{abstract}

\maketitle

\section{Introduction}
\label{sec:intro}

Bistable systems occur throughout the natural sciences and, when such systems are subject to noise, one observes probabilistic transitions between the co-existing metastable states.  Such behavior is found in physical systems such as driven nonlinear mechanical systems \cite{Chan_2007, Aldridge_2005}, magnetization switching in ferromagnetic nanoparticles \cite{Wernsdorfer_1997}, and nonlinear electronic transport systems \cite{Rogozia_2001, Tretiakov_2003, Bomze_2012}.  More broadly, noise-induced dynamics plays a central role in systems of great societal importance such as climate dynamics \cite{Lucarini_2012}, epidemiology \cite{Khasin_PRL_2009}, and pulse propagation in neurons \cite{Izhikevich_2007}.  While there has been substantial recent progress in the theoretical and computational study of these phenomena, they are especially challenging to study experimentally due to the wide range of time scales needed for some of the most interesting measurements. 

Nonlinear electronic transport systems are ideally suited for such experiments because: 1) they can be measured over a very wide dynamic range from sub-nanoseconds to many seconds, and 2) their dynamical properties in the absence of noise are relatively well-understood.  In these systems, bistable behavior is often closely connected with the presence of negative differential resistance (NDR) in the intrinsic current-voltage characteristics, i.e., an increasing applied voltage causes a decrease in electrical current.  NDR is observed in a wide range of electronic transport systems, for example, tunnel diodes \cite{Esaki_1958}, double barrier resonant tunneling structures \cite{Chang_1974}, and molecular bridges \cite{Chen_1999}.  While the microscopic origins of the NDR are system-dependent, the behavior of any NDR element in a circuit displays common and inherently nonlinear features.  Among them is the presence of bistability, i.e., the existence of distinct, co-existing current states for a single applied voltage \cite{Bonilla_2010}.  Furthermore, when several NDR elements are placed in a series array, one often observes striking static and dynamic self-organization effects associated with the development of a spatially non-uniform distribution of electric field or, equivalently, individual element voltages \cite{Xu_2010}.  One example of this behavior is provided by a semiconductor superlattice consisting of a series array of doped quantum wells (e.g., GaAs) separated by potential barriers (e.g., AlAs).  In the case of superlattices with wide barrier layers (called weakly-coupled), one may observe multiple current branches in current-voltage ($I$-$V$) characteristics with the number of branches approximately equal to the number of superlattice periods \cite{Wacker_2002, Bonilla_2005}. These branches are associated with a static, spatially non-uniform electric field configuration in which some periods of the structure are in a low-field domain and the others in a high-field domain \cite{Grahn_1991}.  The observation of current branches typically implies bistability or even multistability in the system, because there are voltage ranges for which the branches overlap.  Current branches and static non-uniform electric field distributions are also observed in other semiconductor systems, including quantum cascade laser structures \cite{Lu_2006}, and multiple quantum well infrared detectors \cite{Schneider_2007}.   

All electronic transport structures are subject to a variety of internal and external noise sources \cite{Buckingham_1985, Gardiner_2009}.  For instance, one source that is always present in superlattices and tunnel diodes is the shot noise associated with the motion of electrons across the tunneling barrier interfaces.  In bistable electronic systems, this leads to the possibility of \textit{noise-induced} switching from one current state to another.  In early work on noise-induced switching in superlattices, Luo et al. \cite{Luo_1998} found a stochastically varying switching time when the sample was subject to a step in voltage from zero to a value near the end of a current branch.  Related measurements investigated the probability distributions of relocation times for positive bias steps from one current branch to the adjacent one, where the voltage after the step (i.e., the final voltage) is close to a discontinuity point in the $I$-$V$ characteristic \cite{Rogozia_2001, Rogozia2002}. The mean value of the switching time was found to increase as the final voltage was reduced to a value approaching the discontinuity, but these measurements were not precise enough to infer the presence of scaling behavior in the mean switching time versus applied voltage.  Nonetheless, the measured switching time distributions were observed to shift from a Gaussian form to a form with exponential tails as the mean switching time increased.  Such measurements are consistent with the picture that switching events close to bifurcation points represent an instance of fluctuation-induced decay of a metastable state.  Experimental measurements of the decay statistics from a metastable state have also been reported for a number of systems including Josephson Junctions \cite{Devoret_1987}, ferromagnetic particles \cite{Victora_1989}, chemical reactions \cite{Dykman_1994}, and periodically driven mechanical systems  \cite{Dykman_2001, Chan_2007}.

The connection between electronic systems that exhibit bistability and Brownian particle dynamics was first suggested many years ago.  In 1962, Landauer published a Fokker-Planck equation describing the electrical current in a bistable tunnel-diode circuit in order to understand the interplay of fluctuation effects and device size \cite{Landauer_1962}.  More recently, a set of Fokker-Planck equations for resonant tunneling structures was derived and studied by Matveev and co-workers \cite{Tretiakov_2003, Tretiakov_2005}.  For devices that possess small \textbf{lateral} size relative to a characteristic length (of order of the lateral charge carrier diffusion length, estimated to be of order 1 - 10 microns for typical doped superlattice structures), the current switching process is directly analogous to a one-dimensional Brownian particle moving in a double-well potential. For devices much larger than this size, the dynamics are analogous to a Brownian particle moving in infinite dimensions with escape possible over a manifold of saddle points; in this case, the transition between current branches begins as a nucleation event in a small lateral sub-region of the quantum well.  This work also predicts scaling of the mean switching time $\langle \tau \rangle$ versus the final voltage $V$ and also surprising dependences of the decay process on the lateral size of the tunneling structures of the general form:

\begin{equation}
\label{eq:Scaling_general}
\ln{\langle \tau \rangle} \propto |V_{TH}-V|^{\alpha},
\end{equation}
where $V_{TH}$ denotes the threshold voltage at which the metastable state loses stability and $\alpha$ denotes the scaling exponent.  For samples with a large diameter, they predicted $\alpha = 1$, while for samples with a small diameter they find $\alpha = 3/2$.  In related work on semiconductor systems, Sch{\"o}ll and co-workers introduced reaction-diffusion models for the lateral dynamics of the front that separates the high and low current density states for double barrier resonant tunneling structures \cite{Rodin_2003} and superlattices \cite{Amann_2005}. 

In fact, $\alpha = 3/2$ is expected when the bistability ends in a generic codimension-1 saddle-node bifurcation \cite{Strogatz_2015} and has been experimentally verified using optically-trapped colloidal particles \cite{Dykman_2003}, electromagnetically trapped ions \cite{Lapidus_1999}, and micromechanical oscillators \cite{Chan_2007}.  However, in all of these systems, verification of scaling laws of the form of Eq. \ref{eq:Scaling_general} with $\alpha = 3/2$ has been limited to one or two decades of time scale.  In the present paper, we demonstrate an experimental system that allows for measurement of switching statistics over a much wider dynamic range.

In the context of electronic transport systems, scaling behavior in mean lifetimes gives crucial information about the robustness and noise sensitivity of metastable states, and this is important for potential applications. For a bistable device intended for memory applications one must operate at voltages such that the probability of \textit{spontaneous} state change is negligibly small.  On the other hand, for applications such as random number generation, one may wish to enhance noise-induced transitions \cite{Li_2013, Alvaro_2014}.  On a more fundamental level, successful measurement of the scaling behavior of mean switching time gives insight into the microscopic dynamics of the switching process. In the current paper, the verification of $\alpha = 3/2$ scaling indicates that the switching process is essentially one-dimensional for the tunnel diodes studied and that non-trivial lateral charge dynamics is not playing a key role, at least for the range of noise intensities reported herein.  

The organization of the paper is as follows.  Section 2 presents a simple and illuminating derivation of the $\alpha = 3/2$ scaling law for an idealized one-dimensional system. Section 3 describes the tunnel diode experimental circuit system and key details of the measurement techniques.  Section 4 describes the experimental method for creating switching time distributions and shows a typical result.  Section 5 describes how the measured mean switching times scale and vary with applied voltage and noise intensity near a saddle-node bifurcation point.  Finally, in Section 6 we present conclusions and briefly discuss some open questions.

\section{Review of scaling law for switching times: the one-dimensional case}
\label{sec:circuit}

\begin{figure}[h]
\includegraphics[width=1.0\columnwidth]{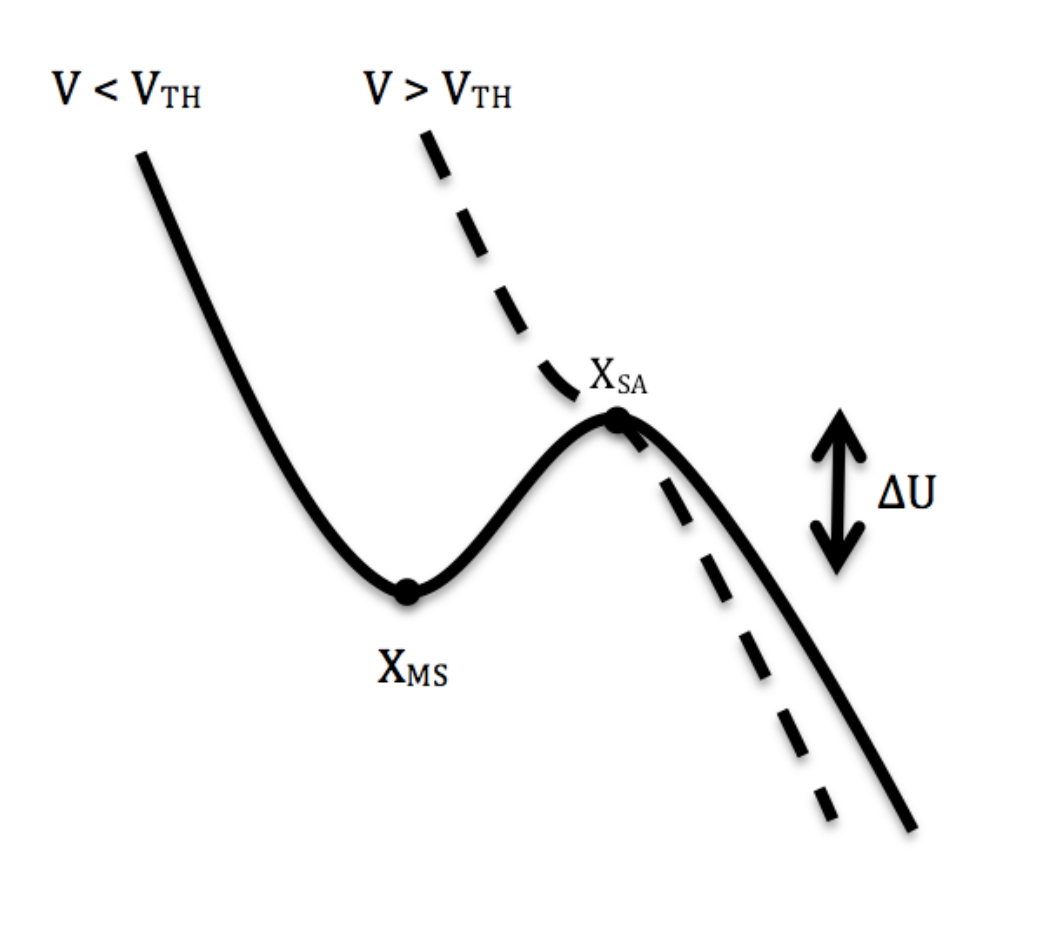}
\caption{Schematic illustration of one-dimensional potential profile just below and just above the threshold for a saddle-node bifurcation.}
\label{fig:1D_escape}
\end{figure}

The $\alpha = 3/2$ exponent for scaling of the escape time due to noise near a saddle-node bifurcation has a remarkable history.   Among the earliest presentations of this behavior must be mentioned the work of Kurkijarvi \cite{Kurkijarvi_1972} (a theoretical study of fluctuations in superconducting Josephson junctions) and, independently, Dykman and Krivoglaz \cite{Dykman_1980} (a theoretical study of fluctuational transitions between co-existing states of a driven nonlinear oscillator).  It is useful here to review a derivation of the basic $\alpha = 3/2$ scaling law based on the simple paradigm of noise-induced escape from a metastable state for an overdamped Brownian particle particle moving in one dimension.  In this case, the equation of motion can be written in Langevin form as:
\begin{equation}
\label{eq:SDE1}
\dot{x}=-U'(x) + \sqrt{2D}w(t),
\end{equation}
where $U(x)$ denotes a potential energy function with nearby metastable and saddle points (e.g., see Fig. \ref{fig:1D_escape}) and $w(t)$ is a unit white noise source, i.e., $\langle w(t)w(t')\rangle= \delta(t-t')$ \cite{Gardiner_2009}. The parameter $D$ denotes the noise intensity; for example, we would have $D = k_B T$ when the system is driven by thermal equilibrium fluctuations.  The coordinate $x$ represents a generic one-dimensional reaction coordinate;  in the context of electronic transport systems, this might correspond to voltage drop across a nonlinear device or the current through it. 

If we assume that the particle starts at the metastable point, then the mean escape time across the potential energy barrier can be determined by the famous Kramers' formula \cite{Kramers_1940, Gardiner_2009}:

\begin{equation}
\label{eq:Kramers}
\langle \tau \rangle \approx \frac{2 \pi}{\sqrt{|U''(x_{MS})U''(x_{SA})|}}\exp{\Big[\frac{\Delta U}{D}\Big]}.
\end{equation}
Here, $\langle \tau \rangle$ denotes the mean escape time, i.e., the mean time for the system to reach the saddle point (SA) due to noisy kicks, given that it started at the metastable point (MS).  The potential energy barrier height is expressed as $\Delta U := U(x_{SA})- U(x_{MS})$, where the potential energy function has a (metastable) minimum at $x_{MS}$ and a saddle point (local maximum) at $x_{SA}$ as depicted in Fig. \ref{fig:1D_escape}.

Equation (\ref{eq:Kramers}) is valid provided that $D$ is not too large, i.e., we should be in the \textit{low noise} limit so that the exponent is much greater than unity.  There should be no intervening saddle points or local minima between metastable and saddle point, i.e., the topology of the potential should be essentially that depicted in Fig. \ref{fig:1D_escape}.  In gradient systems in arbitrary dimension, an analogous expression holds with a prefactor that contains factors corresponding to the components of the Hessian matrix of second derivatives (i.e., frequencies) of potential energy. For \textit{non-gradient} systems (i.e., systems for which the deterministic force field cannot be expresssed as the gradient of a potential energy function) that exhibit a saddle-node bifurcation, the exponential form continues to hold provided that the potential energy is replaced by an appropriate stochastic action \cite{WentzellFreidlin_2012, Heymann_2015}.

We assume that as a system parameter (e.g., applied voltage) is varied a saddle-node bifurcation is approached.  This means that the metastable and saddle points collide and annihilate at a critical or threshold value.  We will call this control parameter $V$ suggestive of applied voltage in an electronic transport experiment.  For simplicity, assume that the saddle-node collision occurs for $V = V_{TH}$ and $x = 0$.  When the control parameter is above the threshold value $V_{TH}$, then the only globally stable fixed point is far away from the origin, and we will assume that this corresponds to some large value of positive $x$.  By contrast, when the control parameter is below the threshold value, there is a stable-unstable pair of fixed points, namely, the metastable and saddle points.

We now give a specific implementation for this family of potential curves.  Generically, we expect that near the saddle-node bifurcation, the potential will have a locally cubic form. The essential idea is that the leading order behavior in a Taylor expansion of a general potential with the same topological features would yield a cubic form.  A suitable family of potential energy functions is given by:

\begin{equation}
\label{eq:1Dpotential}
U(x;V) = K \Big [-\frac{x^3}{3}+A(V)x \Big ],
\end{equation}
where $A(V):=-\beta (V-V_{TH})$, and $K,\beta > 0$.  For this choice of functional form, the potential has a metastable minimum at $x_{MS}=-\sqrt{\beta(V-V_{TH})}$ and a saddle point at $x_{SA}=+\sqrt{\beta(V-V_{TH})}$ provided that $V<V_{TH}$.  At $V=V_{TH}$, the saddle point and metastable annihilate in a saddle-node bifurcation, and for $V>V_{TH}$, there are no equilibrium points.  In the case $V<V_{TH}$, the effective potential energy barrier can be written as $\Delta U = \frac{4}{3}K\beta^{3/2}|V-V_{TH}|^{3/2}$, and we see where the 3/2 scaling exponent arises.  Furthermore, we have $|U''(x)|=|U''(x)|=2K\beta^{1/2}|V-V_{TH}|^{1/2}$ so that we can explicitly evaluate the pre-factor in the Kramer's formula (\ref{eq:Kramers}) and write for the  $V$-dependent mean escape time:

\begin{equation}
\label{eq:mean}
\langle \tau \rangle \approx \frac{ \pi}{2K^2\beta|V-V_{TH}|}\exp{\Big [ \frac{4K\beta^{3/2}}{3D}|V-V_{TH}|^{3/2}} \Big].
\end{equation}
It appears that the mean time diverges very near to the threshold due the diverging pre-factor.  However, it should be noted that this is precisely where the assumption of small noise causes (\ref{eq:Kramers}) to lose validity.  In actuality, the mean escape tends smoothly to zero as we approach the threshold voltage as will be seen in any real experiment.  

Note that the scaling law is \textit{exponential} in the bifurcation parameter $V-V_{TH}$.  For practical purposes, this means that accurate measurement in experimental situations will be challenging, requiring both precise control of the bifurcation parameter (e.g., $V$) and using an apparatus with large dynamic range, i.e., capable of measuring over many decades of time scale.
For exponential scaling laws such as (\ref{eq:mean}) it is useful to have an expression for the logarithm of mean time:
\begin{equation}
\label{eq:logmean}
\log\Big( \frac{\langle \tau \rangle}{\tau_0} \Big) \approx  \frac{4K\beta^{3/2}}{3D}|V-V_{TH}|^{3/2}, 
\end{equation}
where $\tau_0^{-1} := 2K^2\beta|V-V_{TH}|/\pi$.  As already noted, for a  non-gradient, high-dimensional noise-driven system it is generally not possible to calculate the pre-factor $\tau_0$.  However, it may be possible to measure or extrapolate $\tau_0$ in experiments, and this can provide additional insight into the dynamics of the escape process.  It should also be kept in mind that $\tau_0$ may have a qualitatively similar dependence on voltage distance to the bifurcation and serve as a guide for practical fitting approaches.  In particular, one can expect a logarithmic correction to scaling exponent measurements that do not include the possibility of voltage-dependence in the pre-factor.

\section{Experimental circuit and switching time measurement procedure}

\begin{figure}[h]
\includegraphics[width=1.0\columnwidth]{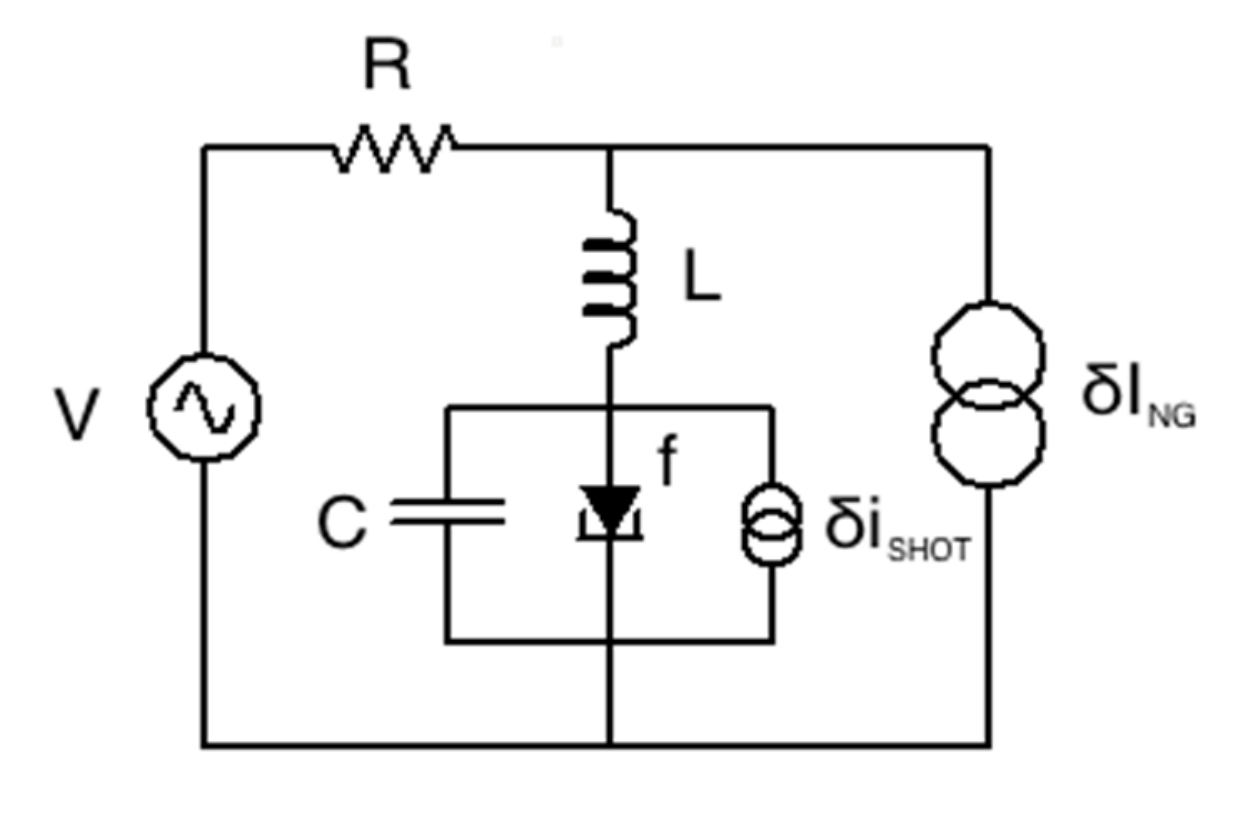}
\caption{Experimental tunnel diode circuit diagram.}
\label{fig:TDCircuit}
\end{figure}

The experimental tunnel diode circuit system is schematically depicted in Figure \ref{fig:TDCircuit}.  The tunnel diode used is model 1N3714 manufactured by Germanium Power Devices Corp.  The physical tunnel diode can be represented in an equivalent circuit diagram as an ideal tunnel diode in parallel with an intrinsic shot noise source and a device capacitance $C$ and also in series with an inductance $L$.  The physical tunnel diode is placed in parallel with: 1) a current source that provides the primary noise signal, and 2) an applied voltage $V$ and effective series resistance $R$.  Parameters appropriate to the current experiment are $R$ = 22 Ohms, $L$ = 0.5 nH, and $C$ = 10 pF.  In order to allow reliable high frequency operation up to the GHz range, the entire circuit is mounted on a patterned copper clad PCB board; connection to measurement devices is made with SMA connectors.

The corresponding circuit dynamical equations can be written in the form:
 
\begin{subequations}
\label{eq:odeTD}
\begin{align}
\dot{v_d}=\frac{I-f(v_d)}{C} -\frac{\delta I_{shot}}{C}, \\
\dot{I}=\frac{V-IR- v_d}{L} -\frac{R\delta I_{NG}}{L},
\end{align}
\end{subequations}
where $v_d$ denotes the voltage across the tunnel diode and $I$ denotes that current flowing through the inductance $L$.  In addition, $\delta I_{shot}$ denotes the intrinsic shot noise of the tunnel diode, $\delta I_{NG}$ denotes the injected signal from the variable noise generator, and $f(v_d)$ denotes the ideal tunnel diode current-voltage characteristic \cite{Okean_1971}. In practice this characteristic curve can be measured directly in the positive differential conductance region and its behavior in the negative differential conductance region can be inferred by fitting measured data to (\ref{eq:odeTD}).  In the experiments reported here, the injected noise is produced by amplifying the intrinsic Johnson-Nyquist thermal noise of a 50 Ohm chip resistor by a series arrangement of three low noise, wide-bandwidth instrumentation amplifiers (Mini-Circuits model ZFL-1000LN+).  

\begin{figure}[h]
\includegraphics[width=1.0\columnwidth]{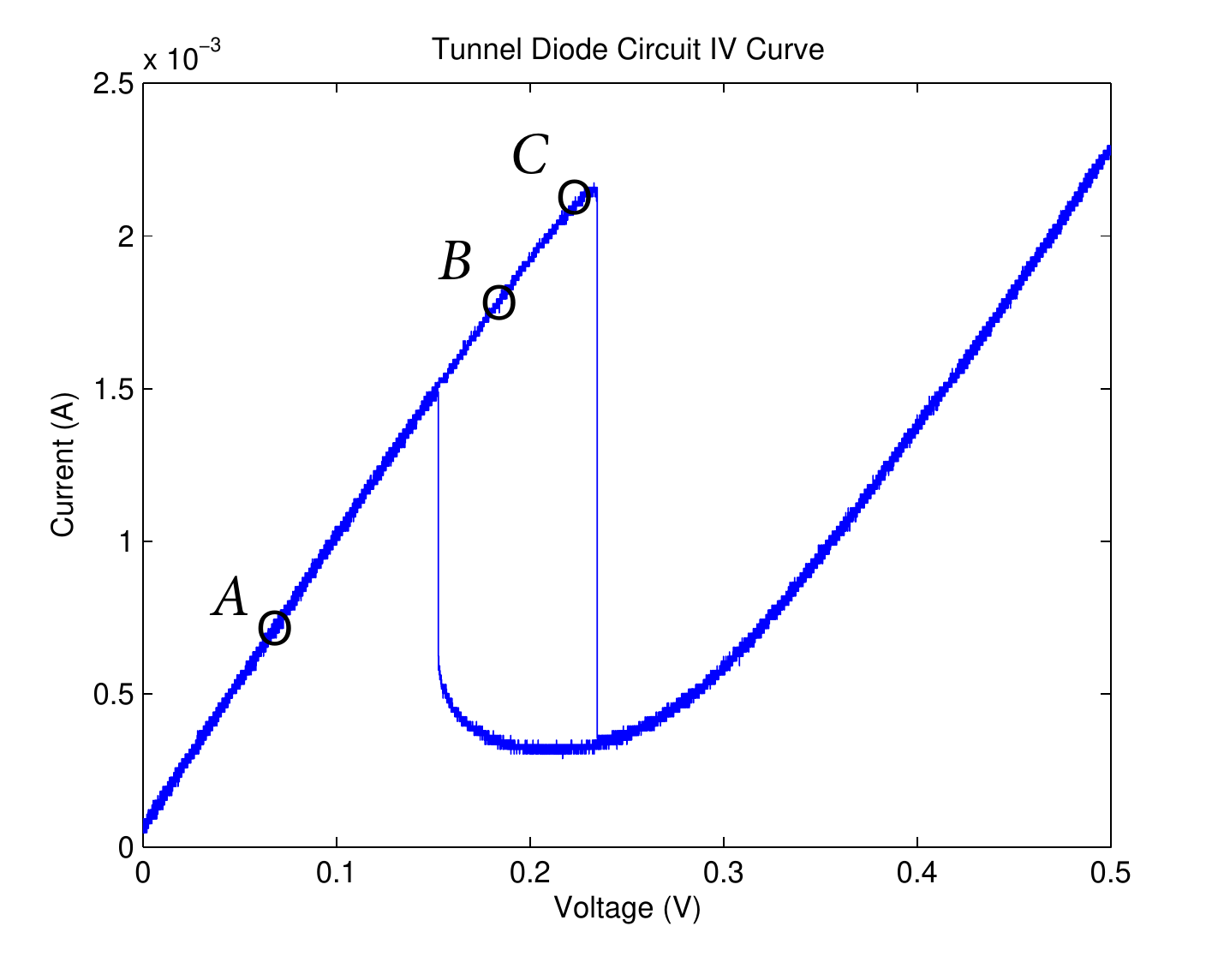}
\caption{Current-voltage curve for the tunnel diode circuit measured for both increasing and decreasing applied voltage and showing hysterisis.  Points labeled $A$, $B$, and $C$ refer to the protocol for switching time measurement.}
\label{fig:IVCurve}
\end{figure}

Figure \ref{fig:IVCurve} shows the measured $I$-$V$ curve of the tunnel diode circuit with no added noise – i.e., no injected signal from the noise generator.  This curve is measured by ramping the applied voltage quasi-statically from 0 V up to 0.5 V and then ramping back down to 0 V at the same ramp speed.  Bistability is clearly evident for the voltage range 0.16 V to 0.23 V.  In the switching measurements described below, we focus on the jump from high to low current state in the upward sweeping direction which occurs at 0.23 V.  It is straightforward to show that this corresponds to a saddle-node bifurcation in the circuit dynamical system (\ref{eq:odeTD}) \cite{Teitsworth_2018}.   Furthermore, we have found that key features of the measured $I$-$V$ curve of Fig. \ref{fig:IVCurve} (e.g., the voltage positions of upward and downward current jumps as well as the overall shape of the curve) are well-described by the model system (7) with appropriate choice of circuit parameters.   

\begin{figure}[h]
\includegraphics[width=1.0\columnwidth]{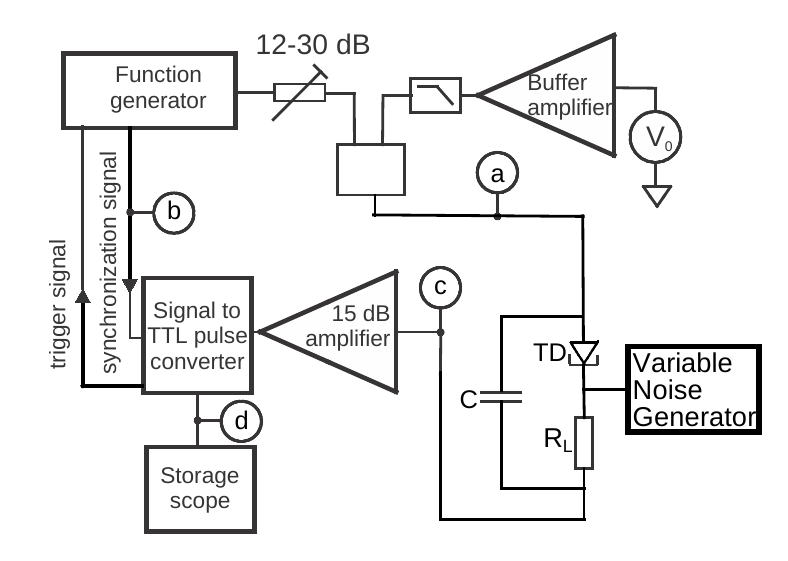}
\caption{Schematic for experimental setup to measure individual switching times and switching time distributions.}
\label{fig:expschematic}
\end{figure}

Next we turn to the experimental method for measuring individual switching times as well as switching time distributions.  Figure \ref{fig:expschematic} shows the measurement circuit which is designed to minimize external interference effects while maintaining high bandwidth.  Key features include attenuating the signal from the digital function generator (Tektronix AFG 3252) by 20 - 30 dB before combining it (in a passive power combiner) with a dc level provided by a precision voltage reference.  This is important since the least significant bit of the function generator output corresponds to approximately 1 mV; without the use of attenuators this would introduce an unintentional and significant noise source.  Another key feature is the use of a signal-to-TTL pulse converter which provides for a self-triggering reset of the function generator to begin another measurement in the series of trials that constitute one ensemble.  This approach substantially cuts the overall run time required to carry out a reasonable number of trials, typically 20000 to 100000 to obtain good statistics.

To prepare the initial state for switching measurements, we utilized a double pulse voltage signal in which the first large pulse takes the system from a globally stable state (e.g., see point marked by $A$ in Fig. \ref{fig:IVCurve}) below the lower current jump at 0.16 V to a metastable state that has low probability to switch (because it is sufficiently far from the upper transition - see point marked by $B$ in Fig. \ref{fig:IVCurve}), and finally it is linearly ramped to the desired metastable state whose statistics we are measuring (see point marked by $C$ in Fig. \ref{fig:IVCurve}).  The use of linear ramp to achieve the final voltage helps maintain the quasi-static nature of the process.  In other words, the system remains close to the time-independent steady state. In a first set of measurements, the time interval between successive measurements was fixed at 7.5 ms.  However, especially for measuring longer mean switching time processes, it is advantageous to employ a self-triggering protocol to allow for the measurement of values that are several orders of magnitude greater than the sampling time (1 ns).  We record up to 100000 switching events for each value of final voltage.  Once the system reaches the saddle point, the measured transition from high to low current state has a duration of about 10 ns; in contrast, the stochastically varying switching times may range up to several seconds and are arbitrarily large for a final voltage farther below the threshold value.  An additional challenge for these measurements is that one cannot directly measure the threshold voltage $V_{TH}$ for which the metastable state ceases to exist in a saddle-node bifurcation.  Instead, voltages are measured relative to an averaged transition value obtained by a quasi-dc $I$-$V$ measurement, yielding an estimated threshold voltage that tends to underestimate the ideal threshold value because the system may escape during the slow voltage ramp before threshold is reached.  

\begin{figure}[h]
    \includegraphics[width=1.0\columnwidth]{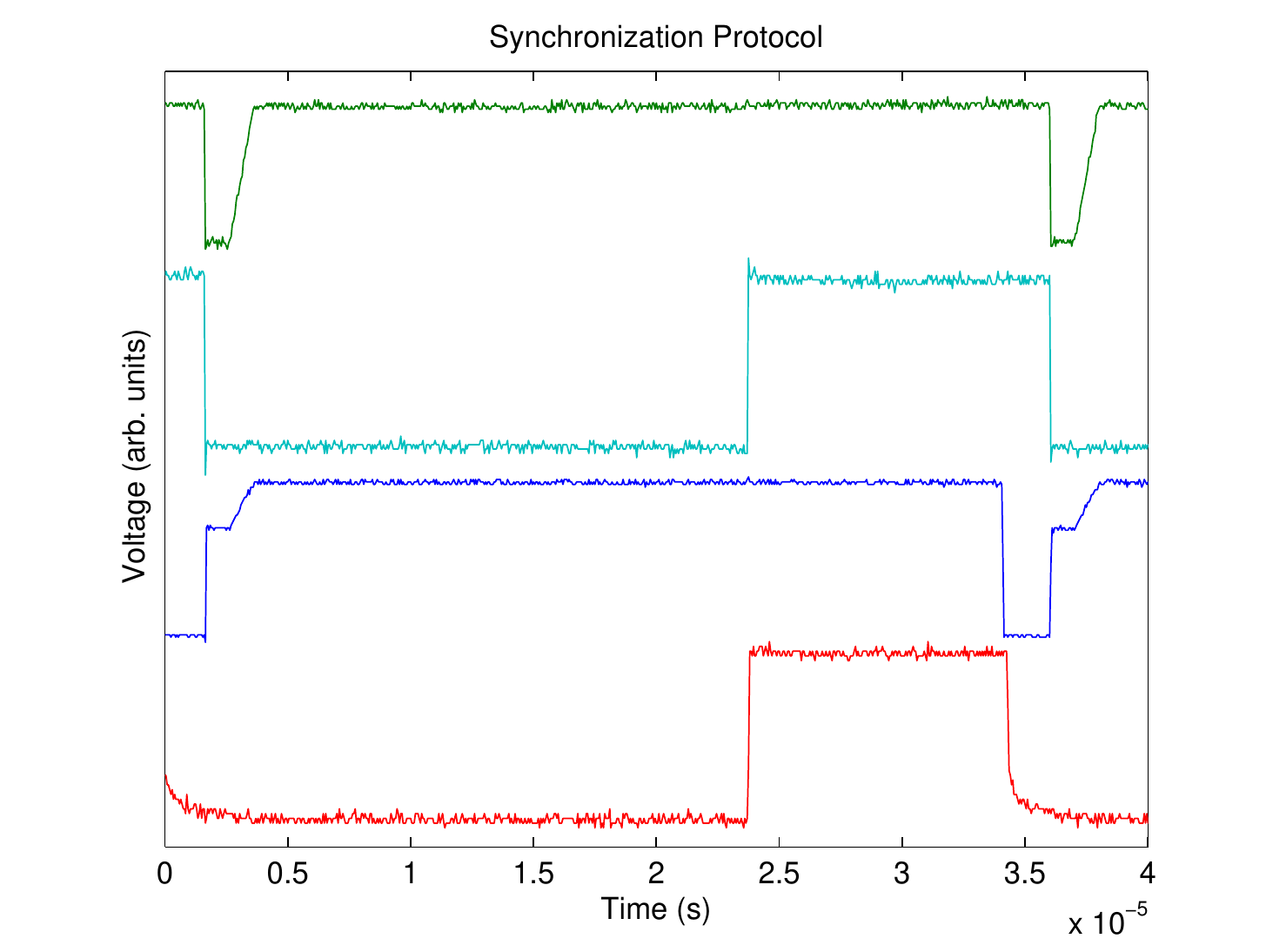}
    \caption{
    Typical experimental time traces of key signals for the measurement of stochastically varying switching times and distributions.  From top to bottom: output voltage signal from function generator; synchronization voltage signal from function generator to indicate the start of a switching time measurement; measured current response of tunnel diode; and digital voltage signal with pulse duration equal to the switching time.}
    \label{fig:Protocol}
\end{figure}

Time traces for key signals in the measurement protocol are shown in Fig. \ref{fig:Protocol}.  Here, the top trace is the output voltage signal of the function generator.  Second from the top is the synchronization voltage signal between function generator and pulse converter that indicates the start of a new switching time measurement (upward pulse) and the start of a new trial (downward pulse) after the current switching event for the previous trial has occurred.  Third from top is the measured current response of the tunnel diode circuit.  Since the current is measured using a current-to-voltage amplifier, this curve is also displayed in voltage units. The bottom trace shows the digital (TTL) voltage pulse-like signal to the storage scope which has duration equal to switching time.  
    
\begin{figure}[h]
    \includegraphics[width=1.0\columnwidth]{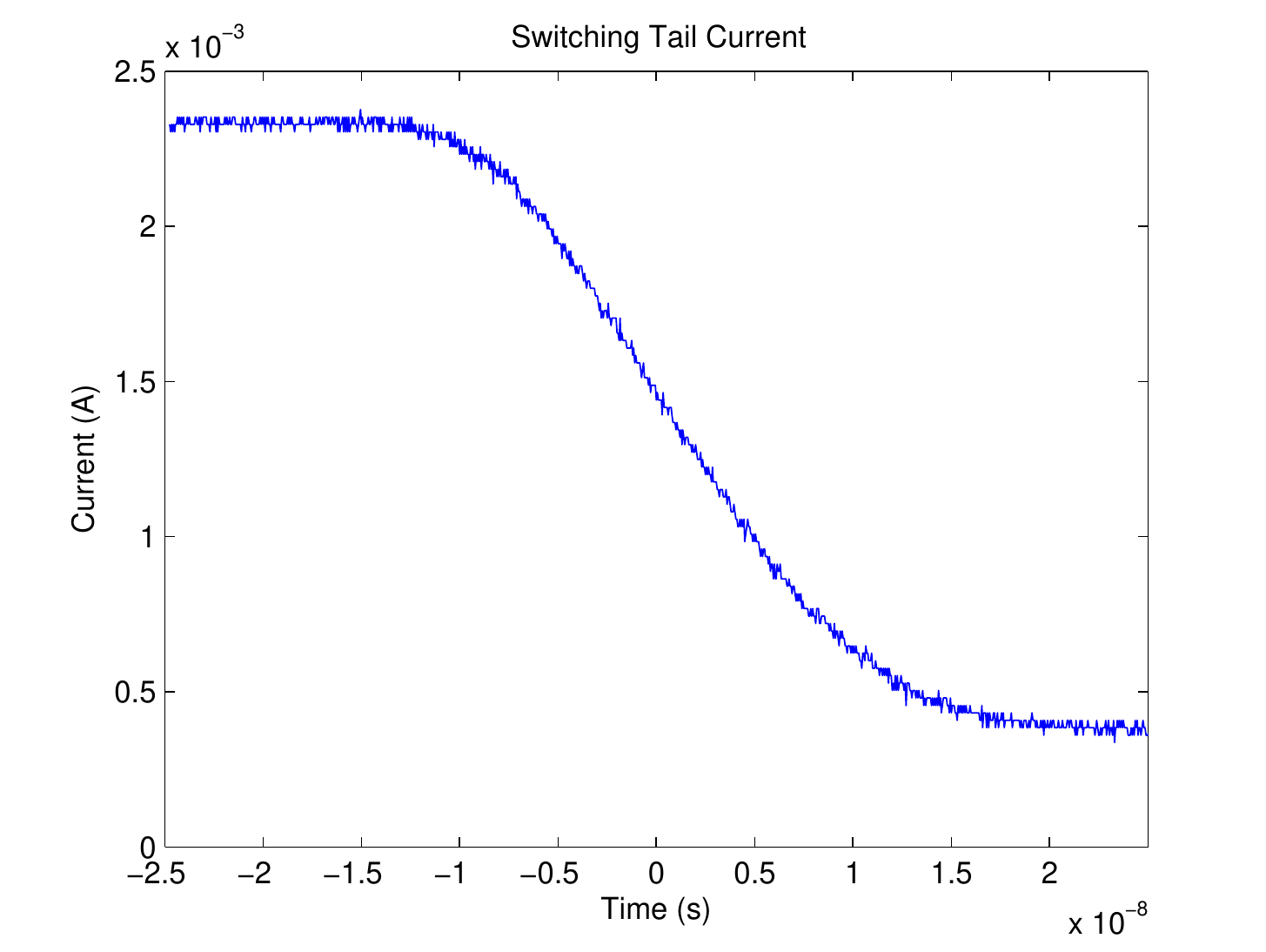}
    \caption{Time-dependent current signal associated with the relatively rapid transition between high and low current states once the saddle point has been reached.}
    \label{fig:tail}
\end{figure}

We also show the relatively rapid transition from high to low current state associated with the switching event in Fig. \ref{fig:tail}.  This signal has a sigmoidal shape and duration of approximately 10 ns, and these features are very similar for different trials (apart from small effects of system noise) and also relatively independent of distance to the threshold voltage $V_{TH}$.  It should be noted that the transition time between the high and low current states in the tunnel diode is much shorter than for analogous measurements in superlattices, which found durations of order 200 ns \cite{Bomze_2012}.

\section{Measurement of switching time distributions}
\label{sec:summary}

\begin{figure}[h]
    \includegraphics[width=1.0\columnwidth]{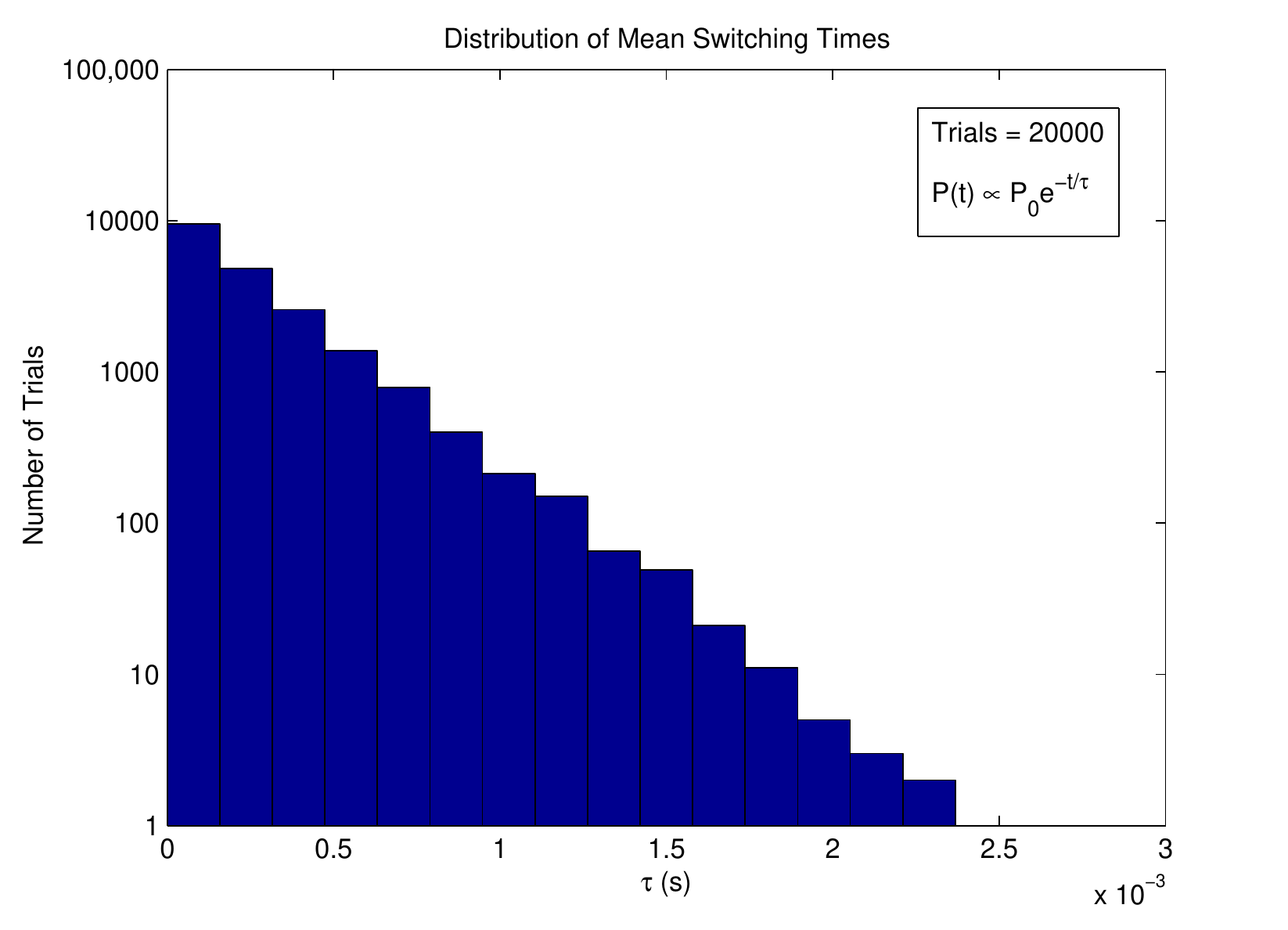}
    \caption{
    Typical measured distribution of stochastically varying switching times for final voltage below the threshold value, i.e., $V < V_{TH}$.}
    \label{fig:Distribution}
\end{figure}

A typical measured histogram of switching time is shown in Fig. \ref{fig:Distribution}. The distribution shows clear exponential decay over a wide range of time scales and is consistent with decay via noise-induced escape over a single saddle point.  The number of trials for the displayed data is 20000 and, in this case, the measured mean switching time is approximately 0.3 ms.  It is also possible to extract a characteristic exponential decay time by fitting the measured distribution to an exponential decay form $P(t)\propto \exp(-t/\tau)$.  Except for final voltages very near the threshold value, the exponential decay time and measured mean decay time agree closely.  As the final voltage approaches the threshold, the measured mean time asymptotes to a value bounded below by the transition time from the high to low current state described in the previous Section, while the extracted exponential time continues to drop towards zero.  It is interesting to note that in similar measurements of switching time distributions carried out for semiconductor superlattices, exponential decay was seen only for an intermediate range of times.  At long times, the distributions unexpectedly transitioned to an algebraic decay, a behavior that is not expected when the noise-induced escape proceeds via a single pathway in phase space.  It was suggested that this feature provides evidence of a fundamentally more complex decay process, for example, the presence of multiple escape pathways that might be associated with structural disorder in the studied devices \cite{Bomze_2012}.  No signs of non-exponential decay in switching time distributions were observed in the measurements reported in this paper, and this suggests that such structural disorder effects are not significant for the noise-driven tunnel diode system.

\section{Mean switching time scaling measurements}
\label{sec:area}

\begin{figure}[h]
    \includegraphics[width=1.0\columnwidth]{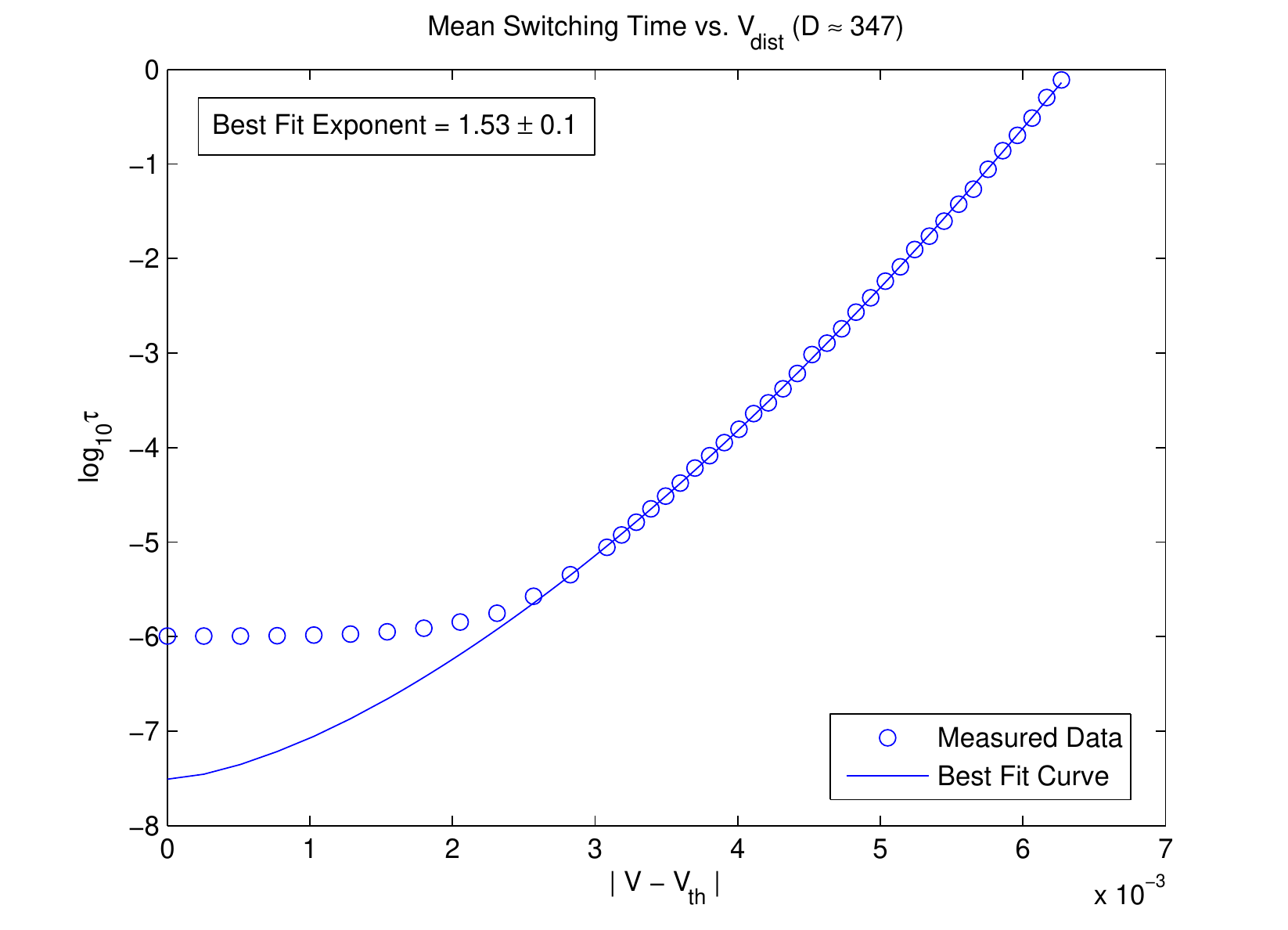}
    \caption{Measured mean switching time versus voltage distance to threshold for a dimensionless noise intensity $D = 347$.}
    \label{fig:Scaling1}
\end{figure}

We now describe the determination of mean switching time from the high to low current state as a function of the final voltage distance to threshold.  For fixed external noise intensity, switching distributions are measured for several final voltages that are just below and in the neighborhood of the threshold voltage determined from the static current-voltage curve, cf. Fig. \ref{fig:IVCurve}.  From each switching time distribution the mean switching time is computed and then plotted vs. the distance between final voltage and threshold voltage, i.e., $V- V_{TH}$. The results are then plotted on a log-linear scale, cf. Fig. \ref{fig:Scaling1}, which shows an example of this procedure for the case where external noise intensity has dimensionless value $D = 347$.  This means that the externally applied noise intensity is 347 times the estimated internal shot noise intensity, $2 e I \delta f \approx 6.7 \times 10^{-13} \text{ V}^2$.  For relatively large voltage differences $V- V_{TH}$ the measured points follow closely a curve that is fitted to an expression similar to (\ref{eq:Scaling_general}) above with best fit exponent determined to be $\alpha = 1.53 \pm 0.10$.  This agrees with the theoretical expectation of $\alpha = 3/2$ to within experimental error.  It is important to note that scaling behavior has been verified over six orders of magnitude in time scale with the shortest mean times on the order of microseconds and the longest on the order of seconds.  At short mean switching times (i.e., when the applied voltage $V$ is close to the bifurcation voltage $V_{TH}$), the measured mean times appear to be independent of voltage distance.  One reason for this that the effective potential energy barrier is small relative to noise intensity and (\ref{eq:Kramers}) is no longer accurate since the exponent approaches zero.

\begin{figure}[h]
    \includegraphics[width=1.0\columnwidth]{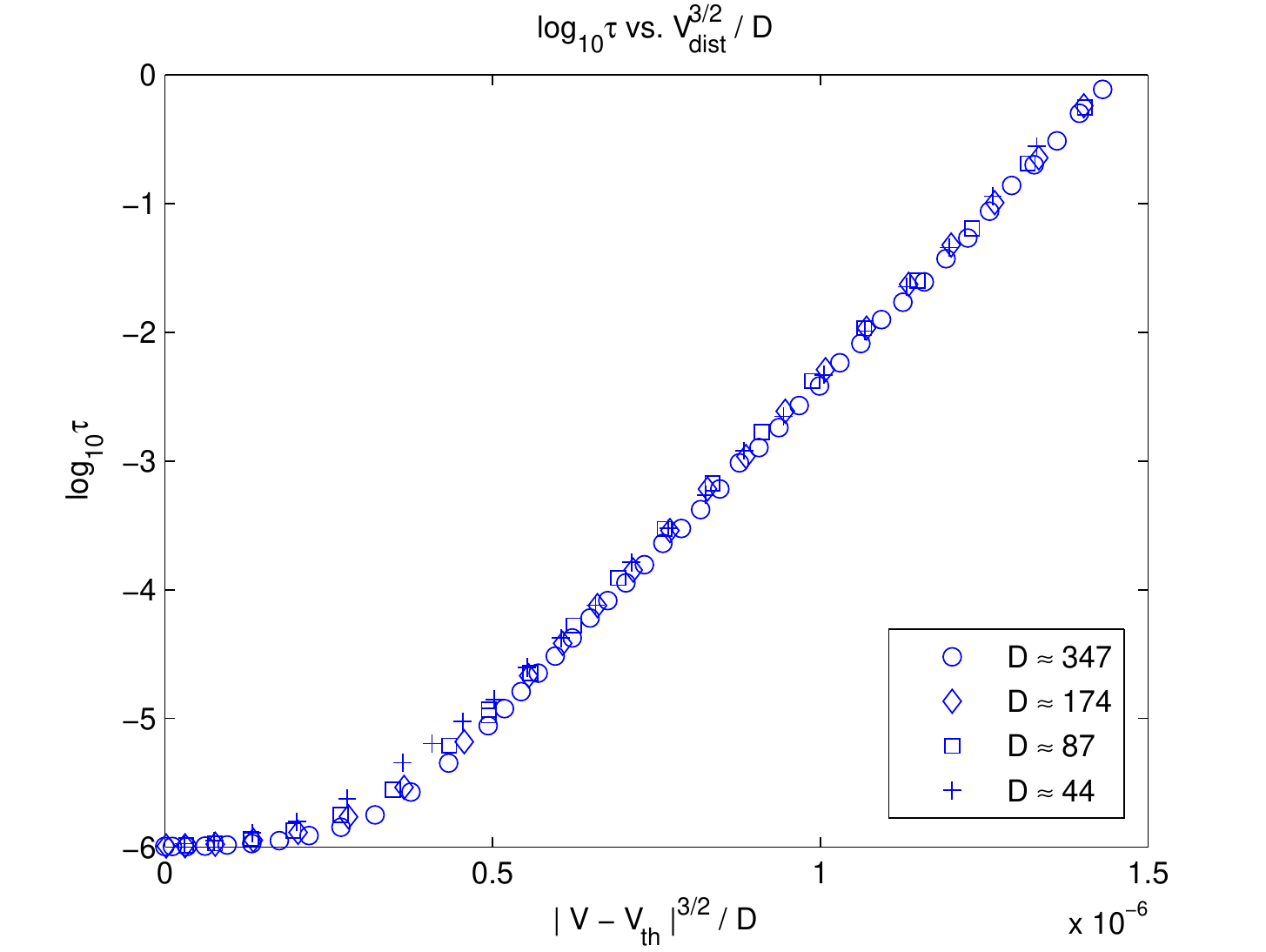}
    \caption{Mean switching time as a function of scaling variable $|V-V_{TH}|^{3/2}/D$ for four different values of relative noise intensity, $D = 44, 87, 174, 347$.}
    \label{fig:Scaling_full}
\end{figure}

In order to explore the robustness of the scaling behavior shown in Fig. \ref{fig:Scaling1}, we have performed similar measurements for a range of different external noise intensities ranging from $D = 1 \rightarrow 1400$.  We find clear evidence of 3/2-power scaling behavior for $D$ in range $40 \rightarrow 400$.  Figure \ref{fig:Scaling_full} plots four sets of data for the noise intensity values $D =$ 44, 87, 174, and 347.  In this case, we plot the logarithm of mean switching time on the vertical axis vs. the quantity $|V-V_{TH}|^{3/2}/D$. The fact that all four of the data sets collapse onto the same curve for larger values of voltage difference indicates the validity and robustness of the theoretical scaling expression of form in (\ref{eq:logmean}).  This supports the idea that, over this range of noise intensities, the switching process is well-described as noise-induced escape over an effective potential barrier in one dimension. 

We have also investigated scaling measurements for larger and smaller noise intensities which we briefly describe now.  For small relative noise intensities below 40, it is difficult to verify scaling behavior because the range of voltages that give measurable switching times is smaller.  In principle, this requires ever more precise control of applied voltage levels which is at odds with the requirement to maintain a large 1 GHz measurement bandwidth.  In our experimental setup, this limits applied voltage resolution to values no smaller than about 10 $\mu$V.  Incidentally, it is for this reason that the scaling regime cannot be accessed when the tunnel diode is driven only by its internal shot noise alone.  Based on our measurements, one can estimate that the voltage range over which the mean time would grow from microseconds to seconds would less than one microvolt which is not experimentally accessible.

On the other hand, for large relative noise intensities in excess of 400, the dependence of mean switching time versus voltage difference can be measured precisely.  In preliminary measurements, we have found evidence for a more complex dependence of mean switching time on voltage distance to the threshold.  In particular, measured dependences suggest that the scaling behavior itself shifts with voltage distance.  At small voltage differences and relatively short mean times, the scaling follows approximately the 3/2 power.  At larger voltage differences, the scaling exponent appears to trend to smaller values closer to $\alpha \approx 1$.  Interestingly, an exponent of $\alpha = 1$ has been predicted by Tretiakov et al. for tunneling structures with relatively large cross sectional areas \cite{Tretiakov_2005}.  Furthermore, in related work, it has been shown that in samples with a quasi-one dimensional strip-like geometry (i.e., where the sample width is small compared to the length) one can obtain scaling behavior in the logarithm of the lifetime of the form $|V-V_{TH}|^{5/4}$ \cite{Tretiakov_2006}.  On the other hand, another possible explanation for a shift in apparent scaling behavior at large voltage distances is the increasing significance of terms beyond the cubic approximation in the effective potential, cf. (\ref{eq:1Dpotential}).  The exploration of this regime requires additional experimental measurements, and it would also be useful to carry out experiments on custom-fabricated tunnel diodes with different cross sectional dimensions and shapes.

\section{Concluding remarks}
\label{sec:summary}

The tunnel diode experimental results described in this paper grew out of a desire to precisely characterize the dependence of metastable lifetimes on system parameters in bistable systems in general, and to look for evidence of scaling behavior in particular. As noted above, previous explorations of related behavior in other experimental systems have often been limited by relatively small dynamic range.  In this paper, we have presented experimental measurements of switching time statistics and scaling of extracted mean switching times versus applied voltage near a saddle-node bifurcation in a tunnel diode circuit system.  A key result of the paper is to verify the $\alpha = 3/2$ scaling law over an exceptionally wide range of time scales and noise intensities.  To our knowledge, these are the first such far-from-equilibrium scaling experiments that incorporate tunnel diodes as a key element.  More generally, tunnel diode circuits have been shown to provide a precise and high-bandwidth experimental platform for exploring bifurcation phenomena both with and without noise.

\section{Acknowledgments}
\label{sec:summary}

The authors thank Dr. Konstantin Matveev for helpful conversations concerning theoretical predictions for scaling laws in tunneling structures. 

\section{Author contributions}
\label{sec:summary}

ST carried out overall experimental design as well as measurement analysis.  MO carried out experimental measurements, data analysis, and figure preparation.  YB carried out circuit design and construction, as well as development of the protocol for measuring switching time distributions.

\end{document}